\documentclass[twocolumn,showpacs,amssymb,prd]{revtex4}

\usepackage{graphicx}
\usepackage{dcolumn}
\usepackage{bm}
\usepackage{epsfig}

\begin{document}
\newcommand{\gsim}{\hbox{\rlap{$^>$}$_\sim$}}
\newcommand{\lsim}{\hbox{\rlap{$^<$}$_\sim$}}

\title{The Smoking Guns of Neutron Stars Mergers}

\author{Shlomo Dado and Arnon Dar}
\affiliation{Physics Department, Technion, Haifa, Israel}

\begin{abstract} 
The short hard gamma ray burst (SHB) 170817A that followed GW170817A, 
the first neutron stars merger (NSM) detected in gravitational waves 
(GWs), has shown beyond doubt that NSMs produce beamed SHBs. Its low 
luminosity and other properties that differ from those of ordinary 
SHBs were predicted by the cannonball model of gamma ray bursts. 
Low luminosity (LL) SHBs are mainly ordinary SHBs viewed far 
off-axis. They are produced mainly by nearby NSMs.  Because of 
beaming,  most of the NSMs, including those within the current 
horizon of  Ligo-Virgo, produce SHBs most of which are invisible 
from Earth. But, their pulsar wind nebula powered by the spin 
down of the remnant neutron star produces an early-time isotropic
afterglow with a universal temporal shape. This smoking gun of NSMs 
is detectable independent of whether the SHB was visible, or was 
invisible from Earth because of being beamed away.
\end{abstract}

\maketitle

\section{Introduction} 
Gamma-ray bursts (GRBs) are brief flashes of gamma rays lasting between
few milliseconds and several hours [1] from extremely energetic cosmic
explosions [2]. They were first detected in 1967 by the USA Vela
satellites. Their discovery was published in 1973 after 15 such events
were detected [3]. Since 1967, the origin and production 
mechanism of GRBs have been among the major puzzles in astrophysics. 

GRBs fall roughly into two classes [4], long duration ones (long 
GRBs) that last more than $\sim$ 2 seconds, and short hard bursts 
(SHBs) that typically last less than 2 seconds. For the first 3 
decades after their discovery, the origin of both types of GRBs were 
completely unknown. This has changed dramatically after the launch of 
the BeppoSAX sattelite in 1996. Its sky localization of GRBs and its 
discovery of their X-ray afterglow [5] led to the discovery of their 
afterglow at longer wavelengths [6,7], which led to the discovery of 
GRB host galaxies [8] and their typical redshifts [9], and  the 
association of {\bf long GRBs} with supernova (SN) explosions of type 
Ic [10]. Following measurement during the past 20 years, mainly with 
the X-ray satellites HETE, Swift, Konus-Wind, Chandra, Integral, 
XMM-Newton, and Fermi, with the Hubble space telescope, and with 
ground based telescopes, provided the detailed properties of the 
prompt and afterglow emissions of GRBs over the entire 
electromagnetic spectrum. They also provided information on their 
host galaxies, their location and environment within their 
hosts, and their production rate as function of redshift.

Progress in identifing the origin and production mechanism of SHBs 
was much slower. Until recently, SHBs were believed [11] to be 
produced in merger of neutron stars [12,13] and in neutron star-black hole 
mergers [14] in compact binaries. This belief was based on indirect 
evidence [12]. Recently, however, SHB170817A [15,16] that followed 1.7s 
after the chirp of the gravitational wave (GW) emission from the 
relatively nearby neutron stars merger (NSM) event GW170817
detected with the Ligo-Virgo GW detectors [17-19], has shown beyond 
doubt that NSMs produce SHBs. SHB170817A
however, appeared to be different from  all SHBs 
observed before it [11]. But, using the Cannonball (CB) model of 
GRBs [20-22], we have shown, as detailed below  that SHB170817A 
was an ordinary SHB  whose  
properties appeared to be different from ordinary SHBs  because 
it was viewed from far off-axis. 

The properties of the far off-axis SHBs and their early afterglow, 
associated with  NSMs detected by the current Ligo-Virgo detectors, 
were predicted [23] before the GW170817-SHB170817A event. In the 
CB model, SHBs are highly beamed.   
NSM events  within the detection horizon of Ligo-Virgo  produce 
mostly far off-axis SHBs. Such far off-axis SHBs, if detected, 
have a much lower luminosity than that of ordinary near axis SHBs.  

The emission of the nebula surrounding the NSMs  that is powered 
by the spin down of the newly born neutron star remnant, is 
isotropic. It dominates the early time afterglow of all SHBs, 
independent on whether the SHB is viewed from near axis or from 
far-off axis [24].  

The late-time beamed afterglow of SHBs is 
usually too faint to be visible, unless the SHB is very nearby, 
such as SHB 170817A. The behavior of such a far off-axis beamed 
afterglow is also well predicted by the CB model of SHBs [25].
 
\section{Universal Properties of SHBs} 
On August 17, 2017, GW170817, the first neutron 
star merger event was detected in gravitational waves by 
Ligo-Virgo [17]. It was followed by a short gamma ray burst, 
SHB170817A, detected by the Fermi [15,26] and Integral [16] satellites, 
$1.74\!\pm\!0.05$ s after the chirp ending the arrival of 
gravitational waves. It was the first indisputable NSM-SHB 
association, indicating that SHBs probably are produced by
NSMs. This is  despite the fact that SHB170817A [26] and its 
{\it late-time} X-ray [27] and radio [28] afterglows look very 
different from SHBs and their X-ray and radio afterglows,
which were observed before [12].

\subsection{Prompt Emission Correlations} 
In the CB model, GRBs 
and SHBs are produced by inverse Compton scattering (ICS) of 
glory photons by a narrow jet of CBs with a large bulk motion 
Lorentz factor $\gamma\!\gg\!1$ launched by a central engine. 
The glory --an ambient light surrounding the launch site-- has 
a cutoff power law (CPL) spectrum, 
$\epsilon\,(dn/d\epsilon)\!\propto\! 
\epsilon^{1-\alpha}exp(-\epsilon/\epsilon_p)$. For a burst at 
redshift $z$ and a viewing angle $\theta^2\!\ll\!1$, and 
$\alpha\!\approx\! 1$ the peak energy of their time-integrated 
energy spectrum satisfies $(1\!+\!z)\, E_p 
\!\propto\!\gamma\,\delta$, and their isotropic-equivalent total 
gamma-ray energy satisfies $E_{iso}\!\propto\!\gamma\delta^3$ 
[29], where the Doppler factor 
$\delta\!=\!1/\gamma(1\!-\!\beta\,cos\theta)$ satisfies to a 
good approximation, $\delta\!\approx\! 
2\gamma/(1+\gamma^2\theta^2)$. Hence, assuming that GRBs 
and SHBs are roughly standard candles in their rest frame, 
GRBs and SHBs that are mostly viewed from an angle 
$\theta\!\approx\!1/\gamma$, satisfy
\begin{equation}
(1+z)\,E_p\!\propto\! [E_{iso}]^{1/2}, 
\end{equation}
while far off-axis ($\theta^2\gg 1/\gamma^2$) SHBs satisfy
\begin{equation}
(1+z)\,E_p\!\propto\! [E_{iso}]^{1/3}.
\end{equation}
The above correlations are well satisfied by GRBs [25]
as shown in Figures 1,2.
\begin{figure}[]
\centering
\epsfig{file=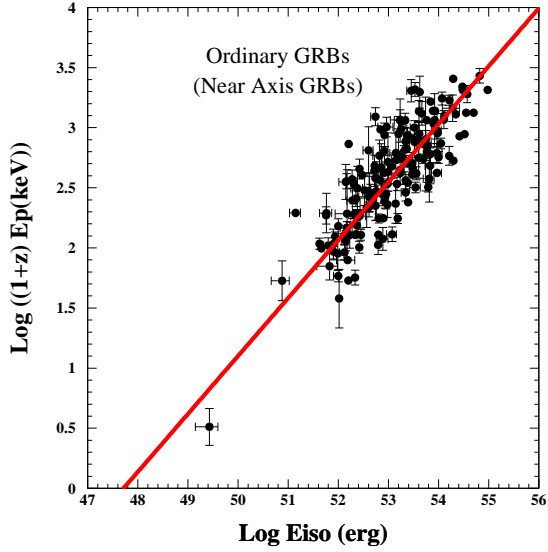,width=8.cm,height=8.cm}
\caption{The peak energy in the GRB rest frame as a function of
the isotropic equivalent total gamma ray
energy of ordinary GRBs viewed near axis. The line is the
correlation predicted by the CB model as given by Eq.(1).}
\end{figure}
\begin{figure}[]
\centering
\epsfig{file=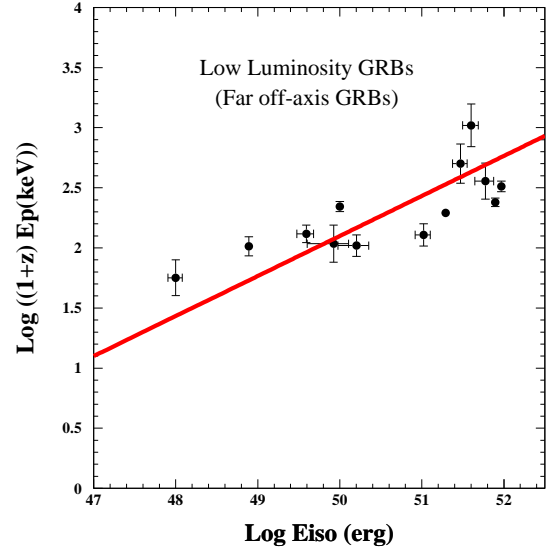,width=8.cm,height=8.cm}
\caption{The $[E_p,E_{iso}]$ correlation in low luminosity
(far off-axis) LGRBs. The line is the CB model prediction
as given by Eq.(2).}
\end{figure}

As shown in Figure 3, SHBs  appear to obey 
Eq.(1), as expected from ordinary (OR) near axis 
($\gamma\theta\!\approx\!1$) SHBs. 
\begin{figure}[]
\centering  
\epsfig{file=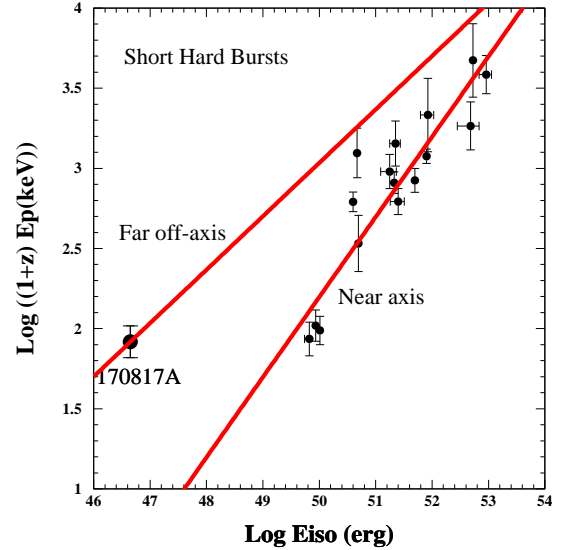,width=8.cm,height=8.cm}
\caption{The $[E_p,E_{iso}]$ correlations in SHBs.
The lines are the CB model predictions
given by Eqs (1) and (2).}
\end{figure}
If, however, 
SHB170817A were an ordinary  standard candle SHB viewed from 
a far off-axis angle $\theta$, then its $E_{iso}$ and $E_p$ should have 
satisfied, 
\begin{equation}
E_{iso}\! \approx\! \langle\! E_{iso}(OR)\!\rangle/
[\gamma^2\,(1\!-\!cos\theta)]^3 \,,
\end{equation}
and
\begin{equation}
(1+z)\,E_p\!\approx\! \langle (1+z)\,E_p(OR)\rangle/
[\gamma^2\,(1\!-\!cos\theta)].
\end{equation}

Indeed, the very small $E_{iso}\!\approx\! 5.6\times 10^{46}$ erg of 
SHB170817A [26]  relative to the mean 
$\langle E_{iso}\rangle\!\approx\!1\times 10^{51}$ erg of OR 
SHBs, as given by Eq.(3), yields $\gamma\theta\!\approx 7.2$.
Assuming the same redshift distribution of GRBs and SHBs with a 
mean value $z\!\approx\!2$, and $\langle\! E_p\!\rangle\!=\!650$
keV [30] yields  $\langle(1\!+\!z)\,E_p\rangle\! \approx\! 1950$ 
keV. Consequently, Eq.(4) with $\gamma\theta\!\approx\! 7.2$ 
yields  $(1\!+\!z)\,E_p\!\approx\!75$ keV for SHB170817A, in 
agreement with the  value $(1\!+\!z)E_p\!=\!82\!\pm\!23$ 
keV ($T_{90}$) reported in [15] and $E_p\!\approx\! 
65\!+\!35/\!-\!14$ keV, estimated in [31]  
from the same data.  This seems to confirm that SHB170817A
was an ordinary SHB viewed from  
a far off-axis. A more direct confirmation is 
provided [32] by the reported apparent superluminal velocity 
of its late-time  radio afterglow [33].

\subsection{Pulse Shape}
SHBs, like long GRBs, are a superposition  of pulses produced by a   
jet of CBs by ICS of glory photons with a CPL spectrum, 
Their time sequence is unpredictable, but their shape is predictable. 
It is given approximately  by [21]. 
\begin{equation}
E{d^2N_\gamma\over dE\,dt}\!\propto\!{t^2\over(t^2\!
            +\!\Delta^2)^2}\,E^{1-\alpha}\,exp(-E/E_p(t)) 
\end{equation} 
where $\Delta$ is approximately the pulse peak time in the 
observer frame, which occurs when the CB becomes transparent to 
its radiation, and $E_p\!\approx\! E_p(t\!=\!\Delta)$.  In 
Eq.(5), the early-time temporal rise like $t^2$ is produced by 
the increasing cross section, $\pi\, R_{CB}^2\!\propto\! t^2$, 
of the fast expanding CB of a radius $R_{CB}$ when the CB is 
still opaque to radiation. When the CB becomes transparent to 
radiation due to its fast expansion, its effective cross section 
for ICS becomes a constant equal to the Thomson cross section times the 
number of electrons in the CB.  That, and the density of the 
glory photons, which far from the launch point decreases like 
$n_g(r)\!\propto\! 1/r^2\!\propto\! t^{-2}$, produce the 
temporal decline like $t^{-2}$. If CBs are launched along the 
axis  of an accretion disk or a torus with a radius 
$R_g$, then glory photons that intercept the CB at a distance 
$r$ from the center have an incident angle $\theta_{inc}$, which 
satisfies $\cos\theta_{inc} \!=\!-\! r/\sqrt{r^2\!+\! R_g^2}$. 
This yields a $t$-dependent peak energy of the photons undergoing 
ICS, $E_p(t)\!=\!E_p(0)(1\!-\!t/\sqrt{t^2\!+\!\tau^2})$ 
with $\tau\!=\!R_g\, (1+z)/\gamma\,\delta\,c$, and 
$E_p\!\approx\! E_p(t\!\approx \!\Delta)$.
  
For GRBs/SHBs  with $\tau\! \gg \!\Delta$, Eq.(5) yields half 
maximum values at $t\!\approx \!0.41\,\Delta$ and 
$t\!=\!2.41\,\Delta$, which yield a full width at half maximum 
$FWHM\!\approx\! 2\,\Delta$, a rise time from half maximum to peak 
value $RT\!=\!0.59\,\Delta$, and a decay time 
from peak count to half peak, $DT\!=\!1.41\,\Delta$. Consequently 
$RT/DT\!\approx\!0.42$ and $RT\approx 0.30\,FWHM$.   
  
In Figure 4  we compare the measured shape [26] of the first prompt
emission pulse of SHB170817A 
and the CB model pulse shape as given by Eq.(5)  with  the best 
fit parameters $\Delta\!=\!0.62$ s and $\tau\!=\!0.57$ s
($\chi^2/dof\!=\!0.95$). The best fit light curve has a maximum 
at $t\!=\!0.43$ s, a half maximum value  at  $t\!=\!0.215$ s
and $t\!=\!0.855$ s, an asymmetry $RT/DT\!=0.50$ and 
$RT/FWHM\!=\!0.34$. 

\begin{figure}[]
\centering
\epsfig{file=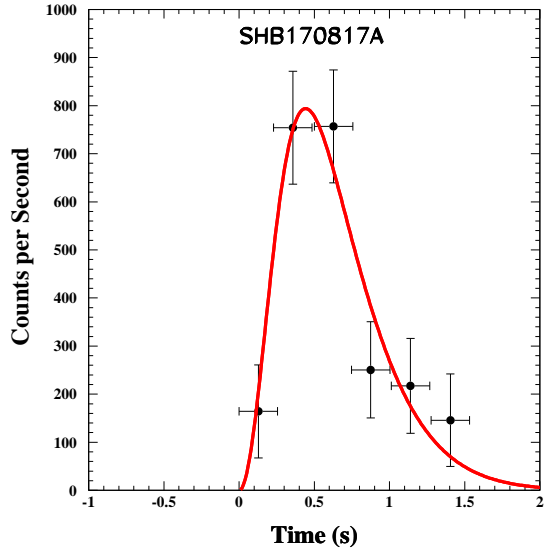,width=8.cm,height=8.cm}
\caption{The pulse shape of SHB170817A measured with the Fermi-GBM
[26]  and the best fit pulse shape given by Eq.(6) with
$\Delta\!=\!0.62$ s and $\tau\!=\!0.57$ s.}
\end{figure}

\subsection{Early-Time Universal Afterglow}
As long as the spin-down of a pulsar
with a period $P(t)$ satisfies $\dot{P}P \!=\!const $,
\begin{equation}
P(t)=P_i(1 +t/t_b)^{1/2},
\end{equation}
where $P_i=P(0)$ is the initial period of the pulsar, $t$ is the 
time after its birth, and $t_b=P_i/2\,\dot{P}_i$. If a constant 
fraction $\eta$ of the rotational energy loss of such pulsar is 
reradiated by the pulsar wind nebula (PWN), then, in a steady state, its 
luminosity 
satisfies $L\!=\!\eta\, I\,w\, \dot{w}$, where $w=2\,\pi/P$ and 
$I$ is the moment of inertia of the neutron star. Hence, in a 
steady state, the luminosity emitted by a PWN satisfies
\begin{equation}
L(t)\!=\!L(0)/(1\!+\!t/t_b)^2.
\end{equation}
Eq.(7) can be written as
\begin{equation}
L(t)/L(0)\!=\!1/(1\!+\!t_s)^2,
\end{equation}
where $t_s\!=\!t/t_b$. Thus, the dimensionless luminosity 
$L(t)/L(0)$ has a simple universal form as function of the 
scaled time $t_s$. For each afterglow of an SN-less GRB (long or 
short) powered by a pulsar, $L(0)$ and $t_b$ can be obtained 
from a best fit of Eq.(7) to the light curve of their measured 
afterglow.

Eqs.(7),(8) are expected to be valid only after the last 
mass accretion 
episode on the newly born pulsar, and after the PWN emission 
powered by the pulsar's power supply has reached a steady state. 
Since the exact times of both are not known, and in order to 
avoid a contribution from the prompt emission, we have fitted 
the observed afterglows of SHBs  with 
Eqs.(7),(8) only well after the fast decline of the prompt emission 
(last pulse/flare or extended emission). This also made 
unimportant the lack of knowledge of the exact time of the beginning of power 
supply by the millisecond pulsar (MSP). The observed early-time 
bolometric afterglow of SHB170817A [34] and its best fit Eq.(7)
are shown in Figure 5.

\begin{figure}[]
\centering
\epsfig{file=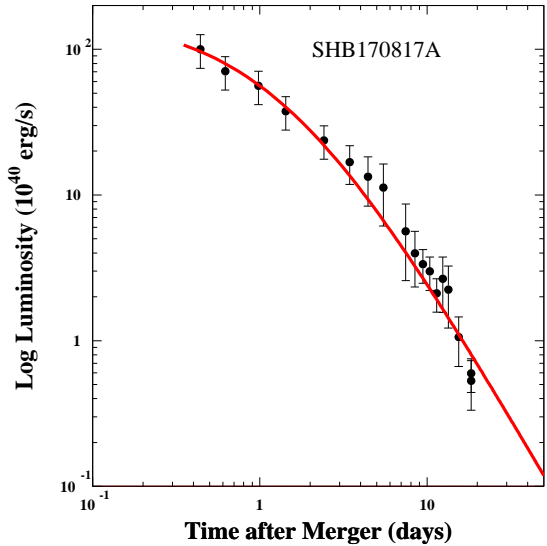,width=8.cm,height=8.cm}
\caption{The best fit CB model lightcurve to the 
observed bolometric lightcurve of SHB170817A [34],
assuming a PWN emission powered by the 
spin down of a neutron star remnant.}
\end{figure}

The initial period of the pulsar enshrouded within a PWN can be 
estimated from its locally measured energy flux $F(0)$
corrected for absorption along the line of sight to the PWN,
its redshift $z$, and its luminosity distance $D_L$,
\begin{equation}
P_i\!=\!{1\over D_L}\,\sqrt{{(1\!+\!z)\,\eta\, \pi\,I \over  2\,F(0)\,t_b}},
\end{equation}  
where $F=L/4\,\pi\, D_L^2$,   
$I\!\approx\!(2/5)\,M\,R^2\!\approx\! 1.12\times 10^{45}\,{\rm g\,cm^2}$,
for a canonical pulsar with  $R\!=\!10$ km and $M\approx 1.4\, M_\odot $,
and $\eta\!<\!1$.
The period derivative can be obtained from the relation
$\dot{P}_i\!=\!P_i/2\,t_b$.   

Eqs,(7),(8) are expected to be valid only after the last accretion 
episode on the newly born pulsar, and after the PWN emission powered 
by the pulsar's power supply has reached a steady state. Since the 
exact times of both are not known, and in order to avoid a 
contribution from the prompt emission, we have fitted the observed 
early time afterglows of SHBs  with Eq.(8) only well after 
the fast decline of the prompt emission (last pulse/flare or extended 
emission). This also made unimportant the lack of knowledge of the 
beginning time of the power supply by the MSP. 

In Figure 6 we plotted 
the  dimensionless X-ray afterglow of 12 SHBs with the  sampled 
afterglows measured with the Swift XRT [35] in the first couple of 
days after burst and the universal behavior given by Eq.(8). The 
values of $L(0)$ and $t_b$ needed to reduce each measured lightcurve 
to the dimensionless form were obtained for each SHB from a best 
fit of Eq.(7) to the observed plateau followed by a fast decline 
phase of the X-ray afterglow.
\begin{figure}[]
\centering
\epsfig{file=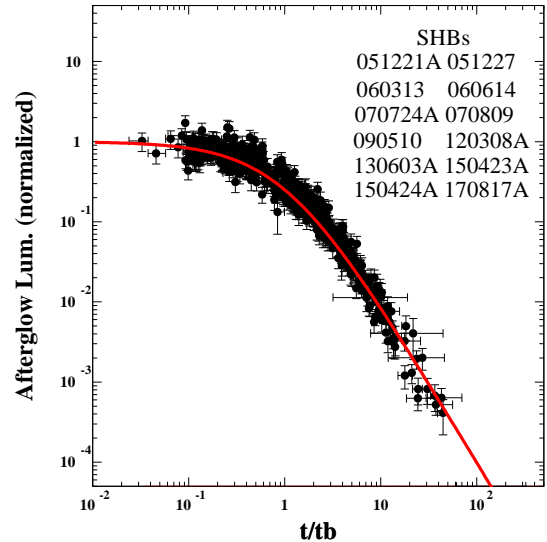,width=8.0cm,height=8.0cm}
\caption{Comparison between the normalized light curve
of the X-ray afterglow of 12 SHBs 
with a well sampled afterglow, measured with the Swift XRT [35]
during the first couple of days after burst 
and the predicted universal behavior given by Eq.(8).
The bolometric light curve of SHB170817A  
is also included.} 
\end{figure}

The values of $P_i$ and $t_b$ for the 11  SHBs with a well 
sampled early time X-ray afterglow are listed in Table I.

\begin{table}[t]
\caption[]{}
\vspace{0.4cm}
\begin{center}
\begin{tabular}{l l l l l l}
\hline
~~SHB~~~& ~~z~~~ & ~$F_X(0)$~ & ~$t_b$ & $\chi^2/dof$ &  $P$  \\
        &      &$[erg/s\,cm^2\,]$& ~[s]~&             &  [ms] \\
\hline
051221A &0.5465& 2.33E-12 &46276 &~~1.04 & 16.2  \\
051227  &0.8   & 1.44E-11 &~2280 &~~0.93 & 19.8  \\
060313  &      & 2.91E-11 &~4482 &~~0.60 &       \\
060614  &0.125 & 1.11E-11 &48931 &~~1.39 & 33.2  \\
070724A &0.457 & 1.18E-12 &18041 &~~1.46 & 43.7  \\
070809 &0.2187 & 3.231E-12&17181 &~~1.05 & 57.6  \\
090510  &0.903~& 3.68E-10 &~~551 &~~1.76 & 7.04  \\
120308A &      & 6.84E-11 &~4742 &~~1.51 &       \\
130603B &0.3564& 6.25E-12 &34382 &~~1.19 & 17.8  \\
150423A &1.394~& 1.31E-11 &~1290 &~~1.36 & 16.0  \\
150424A &0.30  & 5.15E-12 &34461 &~~1.51 & 23.5  \\
170817  &0.0093&          &117374&~~0.60 &       \\
\hline
\end{tabular}
\end{center}
\end{table}
\subsection{The Late-time Afterglow of SHB170817A} 
In the CB model, the observed spectral energy density of the 
unabsorbed synchrotron afterglow produced by a CB is given by 
\begin{equation}
F_{\nu} \propto n^{(\beta_x+1)/2}\,[\gamma(t)]^{3\,\beta_x-1}\,
[\delta(t)]^{\beta_x+3}\, \nu^{-\beta_x}\, ,
\label{Eq10}
\end{equation}
where $n$ is the baryon density of the external medium encountered 
by the CB at a time $t$ and  $\beta_x$ is the spectral index 
of the emitted X-rays, $E\,dn_x/dE\propto E^{-\beta_x}$.
For a constant density, the deceleration of the CB yields a late-time 
$\gamma(t)\propto t^{-1/4}$ and as long as $\gamma^2\!\gg\!1$,
$\beta\!\approx\! 1$ and $\delta(t)\!\propto\!t^{-1/4}/(1-cos\theta)$.  
Consequently, the apparent superluminal velocity of a far off-axis
CB in a constant low-density environment, stays constant, as
long as $\beta\!\approx\!1$, while its late-time spectral
luminosity increases like $t^{(1-\beta_\nu/2)}$. When the CB 
exits  the disk into the halo, it turns into a fast decay  
$\propto\![n(r)]^{(1\!+\!\beta_\nu)/2}$. Approximating the  
disk density perpendicular to the disk by 
$n(h)\!=\!2\,n(0)/(1\!+\!exp(h/d))$
where $d$ is the "skin depth" of the disk, the lightcurve 
of the  afterglow of  SHB170817A can be approximated by, 
\begin{equation}
F_\nu(t)\!\propto\!{(t/t_e)^{1\!-\!\beta_\nu/2}\, 
\nu^{-\beta_\nu}\over
[1+exp[(t\!-\!t_e)/w]]^{(1\!+\!\beta_\nu)/2}} 
\end{equation}
where $t_e$ is roughly the escape time of the CB from the 
galactic disk into the halo after its launch. Eq.(11) is  
compared to the observed late-time X-ray [27] and radio [28] 
afterglows of SHB170817A in Figures 7 and 8, respectively.

\begin{figure}[]
\centering
\epsfig{file=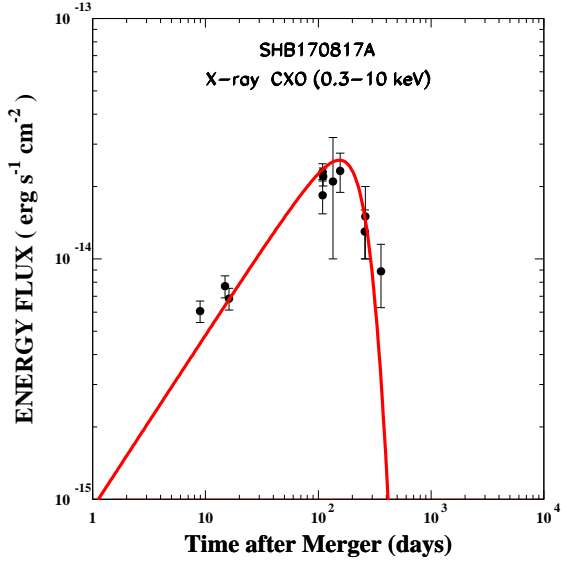,width=8.cm,height=8.cm}
\caption{The lightcurve of the X-ray
afterglow of SHB170817A  measured with the CXO [27]
and the lightcurve predicted by eq.(11) for $\beta_X=0.56$ ,
and the best fit parameters $t_e\!=\!245.6$ d  and $w=63.4$ d.}
\end{figure}

\begin{figure}[]
\centering
\epsfig{file=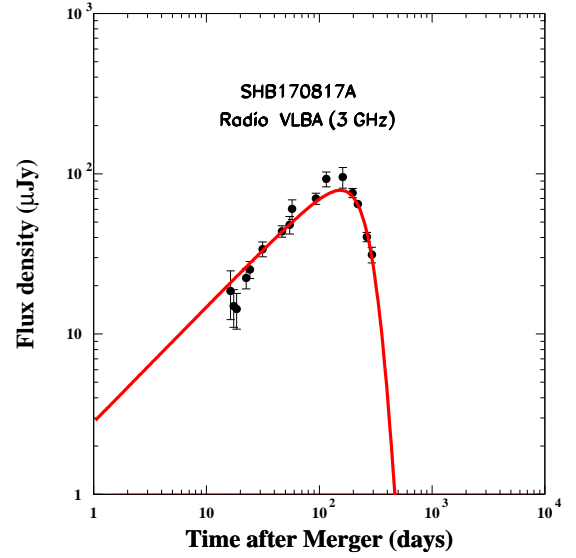,width=8.cm,height=8.cm}
\caption{Left: The measured [28]
lightcurve of the 3 GHz radio afterglow of SHB170817A
and the lightcurve predicted by the CB mode $\beta_r=0.56$,
$t_e\!=\!245.6$ d  and $w=63.4$ d.}
\end{figure}

\subsection{Superluminal Radio Afterglow of SHB170817A }
The VLA and VLBI localization of the source of the late time radio afterglow  
of SHB170817A [33] provided the first successful measurement of the apparent 
superluminal motion of the highly relativistic CBs, which produce the prompt 
emission and  beamed  afterglow of GRBs and SHBs
(two decades after the discovery of the afterglow of GRBs at lower 
frequencies [5-7]). 

The apparent  velocity in the plane of the sky of a highly
relativistic compact (unresolved) source moving at a small redshift
($1\!+\!z\!\approx\!1$) with a {\bf constant} bulk motion Lorentz
factor $\gamma\!\gg\!1$ viewed from an angle $\theta$ relative
to its direction of motion, is given by
\begin{equation}
V_{app}\!=\!\beta\,\gamma\,\delta\,c\, sin\theta/(1\!+\!z)
\!\approx\! {c\, sin\theta\over (1\!+\!z)(1\!-\!cos\theta)}\,,
\end{equation}
which depends only on $\theta$. In that case, 
the angular distance to the afterglow source 
$D_A\!\approx V_{app}\,\Delta t/(1\!+\!z)\,\Delta\theta_s $,
and the angular change $\Delta\theta_s$ in the sky location 
of the source yield
\begin{equation}
{sin\theta \over (1\!-\!cos\theta)} \!\approx\!
        {\Delta\theta_s\, D_A\,(1\!+\!z)\over c\, \Delta t} 
\end{equation} 
In the case of SHB170817A in NGC 4993 at redshift    
$z\!=\! 0.009783$ [36], the angular location of the radio source in the 
plane of the sky has changed during $\Delta t$=155 days (between day 75 
and day 230) by $\Delta\theta_s\!=\!2.68\!\pm\!0.3$ mas  [33].
Thus, assuming the local value of 
$H_0$, $73.52\!\pm\! 
1.62\,{\rm km/s\, Mpc}$ [37], i.e., $D_A\!\approx\! 39.43$  Mpc, 
Eq.(12) yields  $\theta\approx 28\!\pm\! 2$ deg 
($V_{app}\!\approx\! (4.0\pm\! 0.4)c$).
This value of $\theta$ is in agreement with the value $25\!\pm\! 4$ 
deg, which was obtained [38] from GW170817 and its
electromagnetic localization [15,26] assuming the local value  of 
$H_0$ obtained [37] from Type Ia supernovae (SNeIa).

\section{conclusions}
Ordinary SHBs are visible up to very large cosmological distances.
They are produced mostly by neutron star mergers whose rotational axis 
points near the direction of Earth. Low luminosity SHBs are 
produced by nearby neutron star mergers whose rotational axis points 
far off the direction of Earth. Only a small fraction of NSMs  
within the current detection horizon of Ligo-Virgo produce 
SHBS visible from Earth. However, all mergers that end with a pulsating 
neutron star remnant  and a visible or an invisible SHB  produce {\bf an 
early-time isotropic afterglow -the smoking gun of NSMs}. 
These early time afterglows  are  powered by the spin-down of the 
remnant neutron star. They have a universal temporal behavior
common to ordinary SHBs, far off-axis SHBs, and invisible 
(very  far off-axis SHBs), and are visible up to very large 
cosmological distances

\end{document}